# SU(3) Ghosts with Spin

Scott Chapman
Sunnyvale, CA

June 14, 2007

**Abstract**

A new Lorentz-covariant gauge is presented for SU(3). In this gauge, both the ghosts and the gauge fields in the (4, 5, 6, 7) gauge directions acquire half-integral spin. As a result, the ghosts in these directions have the "correct" relationship between spin and statistics, while the gauge fields have the "incorrect" relationship. Consequently, asymptotic ghost states are not forbidden in this gauge and can possibly form new matter states. Conversely, asymptotic gauge fields in the (4, 5, 6, 7) directions are forbidden in this gauge, so the SU(3) symmetry is broken down to an SU(2)xU(1) symmetry asymptotically.

**Introduction**

To quantize SU(3), one must first choose eight gauge conditions – one for each of the eight adjoint gauge directions. The standard choice is to impose the eight conditions $\partial^\mu A_\mu^a = 0$, where $A_\mu^a$ is the gauge field and the index *a* runs from 1 to 8. In this paper, the standard gauge condition will be employed for the (1, 2, 3, 8) gauge directions, but a generalized axial gauge condition of the form $(\Gamma^\mu)^{ab} A_\mu^b = 0$ will be employed for the (4, 5, 6, 7) directions. Here $(\Gamma^\mu)^{ab}$ are constant matrices in the adjoint gauge space that are formed from "spin" matrices $(T^\mu)^{ab}$ that transform as a 4-vector in the presence of global gauge transformations.

The antifield formalism is used in the paper to derive gauge-fixed actions, and the paper begins with a definition of notation for that formalism. Next, the matrices $(T^\mu)^{ab}$ are identified and shown to transform like a 4-vector as long as every Lorentz transformation is accompanied by a corresponding global gauge transformation. The new Lorentz-invariant gauge-fixed action is then presented, and it is shown that four of the ghosts have spin ½, three have spin 1, and one has spin 0. Furthermore, the gauge fields associated with the spin ½ ghosts also acquire half-integral spin due to the global gauge transformation. The paper ends with discussion, including speculative comparisons between the SU(2)xU(1) of this model and that of the Standard Model.

**Antifield Formalism Notation**

The antifield formalism is a Lagrangian-based technique that can be used to create Lorentz-invariant, gauge-fixed actions. In this formalism, the gauge-fixed action for SU(3) is given by [1]

$$S^* = \int d^4x \left\{ -\tfrac{1}{4} F^a_{\mu\nu} F^{\mu\nu a} + A^{*\mu a} D^{ab}_\mu c^b - \tfrac{1}{2} f^{abc} c^{*a} c^b c^c + b^a \bar{c}^{*a} \right\}, \qquad (1)$$

where $F^a_{\mu\nu}$ is the usual field-strength tensor, $c^a$ are ghosts, $b^a$ are multipliers used to enforce gauge conditions, and $A^{*\mu a}$, $c^{*a}$ and $\bar{c}^{*a}$ are "antifields". The "antifields" can be expressed in terms of standard fields through the following relations

$$A^{*\mu a} = \frac{\delta \Psi}{\delta A^a_\mu} \qquad \bar{c}^{*a} = \frac{\delta \Psi}{\delta \bar{c}^a} \qquad c^{*a} = \frac{\delta \Psi}{\delta c^a}, \qquad (2)$$

where $\Psi$ is a "gauge-fixing fermion" that one may freely choose.

To illustrate how the antifield formalism works, it is helpful to use it to derive the standard Lorentz gauges. For standard Lorentz-covariant gauges like $\partial^\mu A^a_\mu = f(x)$, one normally chooses a gauge-fixing fermion of the form

$$\Psi_0 = \int d^4x \left\{ \bar{c}^a \left( \partial^\mu A^a_\mu + \tfrac{1}{2} \xi b^a \right) \right\}. \qquad (3)$$

After using (2) to replace the antifields with standard fields and then integrating $\exp(iS^*)$ over $b^a$, the gauge-fixed action becomes

$$S_0 = \int d^4x \left\{ -\tfrac{1}{4} F^a_{\mu\nu} F^{\mu\nu a} + \bar{c}^a \partial^\mu D^{ab}_\mu c^b - \frac{1}{2\xi} \left( \partial^\mu A^a_\mu \right)^2 \right\}. \qquad (4)$$

This is the standard Lorentz-covariant action with the usual Fadeev-Popov ghosts and gauge-fixing term.

**Construction of Spin Matrices and Transformations**

For the new gauge, one must first find a set of matrices $\left(T^i\right)^{ab}$ that close in an SU(2) "spin" group, but "live" in the 8-dimensional adjoint space of SU(3). The adjoint representation matrices

$$\left(F^a\right)^{bc} \equiv -if^{abc} \qquad (5)$$

"live" in the correct space, and the first three of them close in an SU(2) group

$$\left[F^i, F^j\right] = i\varepsilon^{ijk} F^k \qquad \text{for } i \in (1,2,3), \qquad (6)$$

so they have the right properties. The 8x8 matrices $F^i$ are block diagonal with different blocks having different SU(2) representations: The first three indices form a 3x3 spin 1 representation, indices 4-7 form a 4x4 spin ½ representation, and index 8 forms a 1x1 spin 0 representation.

It is useful to define notation that separates the spin ½ block of $F^i$ from the other blocks:

$$\left(F^i\right)^{ab} = \begin{pmatrix} \left(t^i\right)^{AB} & 0 \\ 0 & \tfrac{1}{2}\left(\tau^i\right)^{\alpha\beta} \end{pmatrix}, \quad \text{where} \quad \begin{matrix} A, B \in (1,2,3,8) \\ \alpha, \beta \in (4,5,6,7) \end{matrix}. \tag{7}$$

The 4x4 matrices $\tau^i$ have the same commutation relations, anti-commutation relations, and eigenvalues as Pauli matrices. As a result, they have the correct properties to act as spin matrices for ghosts in the new gauge. To create Lorentz-covariant actions, one must also define a "time" component:

$$\tau^\mu = \left(1, \tau^i\right). \tag{8}$$

The 4x4 matrices in (8) can be used to build the full 8x8 matrices referred to in the Introduction:

$$\left(T^\mu\right)^{ab} \equiv \begin{pmatrix} (0)^{AB} & 0 \\ 0 & \left(\tau^\mu\right)^{\alpha\beta} \end{pmatrix}. \tag{9}$$

The next step is to find global gauge transformations that will cause the above quantities to transform like 4-vectors.

In analogy with transformations for standard spin ½ fields, one may define the following transformation matrices (see for example [2]):

$$\Lambda_L \equiv \exp\left((i\vec{\omega} + \vec{v}) \cdot \vec{F}\right)$$
$$\Lambda_R \equiv \exp\left((i\vec{\omega} - \vec{v}) \cdot \vec{F}\right) \tag{10}$$

where $\omega^i$ and $v^i$ are real constants representing rotations and boosts, respectively. It will be stipulated here that every time one performs a Lorentz transformation, one must also perform the following corresponding generalized global gauge transformations:

$$A_\mu \equiv A_\mu^a F^a \to \Lambda_L A'_\mu \Lambda_R^\dagger$$

$$c \equiv c^a F^a \to \Lambda_L c \Lambda_R^\dagger$$

$$\overline{C} \equiv \overline{c}^A F^A \to \Lambda_L \overline{C} \Lambda_R^\dagger \qquad \text{indices } A \in (1,2,3,8)$$

$$\overline{c} \equiv \overline{c}^\alpha F^\alpha \to \Lambda_L \overline{c} \Lambda_L^\dagger \qquad \text{indices } \alpha \in (4,5,6,7) \tag{11}$$

In the first equation above, the prime on the right-hand side is a reminder that the Lorentz transformation also has an effect on the spacetime indices of the gauge field. In analogy with standard spinor analyses, these transformations cause quantities like $\text{Tr}(\bar{c}T^\mu c)$ to transform like Lorentz 4-vectors.

Since these global gauge transformations are correlated with Lorentz transformations, they impart spin to the fields. The spin may be calculated by considering infinitesimal spatial rotations then determining the amount that the gauge transformations contribute to the conserved angular momentum. For an infinitesimal rotation, the ghost field transforms as follows

$$c^a \to c^a + \omega^i f^{iab} c_b. \tag{12}$$

Following standard spin derivations (see for example [3]), one finds that the spins of the ghost fields in the *i* direction are given by the eight eigenvalues of $f^{iab}$. In other words, the ghosts in the (1, 2, 3) directions form a spin 1 triplet, the ghost in the 8 direction is a spin 0 singlet, and the ghosts in the (4, 5, 6, 7) directions form a spin ½ quadruplet. These latter four ghosts are particularly interesting since they have the correct relationship between spin and statistics, so they could exist in asymptotic states.

The adjoint indices of the gauge fields undergo the same transformation as the ghosts, so they also acquire new spin in addition to the spin 1 that comes from their spacetime indices. Adding these contributions, the gauge fields in the (1, 2, 3) directions have spin 0, 1, or 2, the gauge fields in the 8 direction have spin 1, and those in the (4, 5, 6, 7) directions have spin ½ or 3/2. The fields in the latter four directions have the wrong relationship between spin and statistics, so they cannot exist in asymptotic states. Therefore, the transformations (11) have the effect of removing those guage fields asymptotically from the theory and thereby breaking the SU(3) symmetry down to SU(2)xU(1) asymptotically.

**A new Lorentz-invariant gauge-fixed action**

It is now possible to write down the following 8 gauge conditions for SU(3):

$$\partial^\mu A_\mu^A = -\tfrac{1}{2}\xi b^A \qquad \text{for indices } A \in (1,2,3,8) \tag{13}$$

$$\text{Tr}(F^\alpha T^\mu A_\mu) = 0 \qquad \text{for indices } \alpha \in (4,5,6,7). \tag{14}$$

Condition (13) is manifestly Lorentz-invariant, even in the presence of generalized global gauge transformations. Given the definitions and transformations of (11), condition (14) is also Lorentz-invariant. This can be seen in the following way: Upon a Lorentz transformation (with corresponding gauge transformation),

$$T^\mu A_\mu \to \Lambda_R \Lambda_L^\dagger T^\mu \Lambda_L A'_\mu \Lambda_R^\dagger = \Lambda_R T^\mu A_\mu \Lambda_R^\dagger, \tag{15}$$

where $\Lambda_R \Lambda_L^\dagger = 1$ has been used. Consequently,

$$\text{Tr}(F^\alpha T^\mu A_\mu) \to \text{Tr}(\Lambda_R^\dagger F^\alpha \Lambda_R T^\mu A_\mu) = \sum_\beta k^{\alpha\beta} \text{Tr}(F^\beta T^\mu A_\mu). \tag{16}$$

The last term is another way to say that $\Lambda_R^\dagger F^\alpha \Lambda_R$ only produces terms proportional to $F^\beta$, where both $\alpha, \beta \in (4,5,6,7)$. As a result, if (14) holds in one reference frame, then it holds in all reference frames accessible from a Lorentz transformation. Equation (14) can also be written in the form $(\Gamma^\mu)^{\alpha\beta} A_\mu^\beta = 0$ with $(\Gamma^\mu)^{\alpha\beta} = \text{Tr}(F^\alpha T^\mu F^\beta)$, which is the form mentioned in the Introduction. In this form it is clear that (14) is a Lorentz-invariant generalization of the axial gauge.

To implement the new gauge conditions in the action, one can define the following gauge-fixing fermion:

$$\Psi_1 = \tfrac{1}{3}\text{Tr}\int d^4x \left\{ \overline{C}(\partial^\mu A_\mu + \tfrac{1}{2}\xi B) + 2(\overline{c}T^\mu A_\mu) \right\}, \tag{17}$$

where the factor of $\tfrac{1}{3}$ compensates for the fact that $\text{Tr}\{F^a F^b\} = 3\delta^{ab}$. Using equation (2) to replace the antifields and then integrating over $b^A$, one finds:

$$S_1 = \int d^4x \left( -\tfrac{1}{4} F_{\mu\nu}^a F^{\mu\nu a} - \frac{1}{2\xi}(\partial^\mu A_\mu^A)^2 + \tfrac{1}{3}\text{Tr}\left\{ \overline{C}\partial^\mu \left( \partial_\mu c - ig[A_\mu, c] \right) \right\} \right.$$
$$\left. + \tfrac{2}{3}\text{Tr}\left\{ bT^\mu A_\mu + \overline{c}T^\mu \left( \partial_\mu c - ig[A_\mu, c] \right) \right\} \right) \tag{18}$$

The first line reflects the standard Lorentz gauge fixing for directions (1, 2, 3, 8), while the second line implements the new gauge fixing for the other directions. Using the transformations of (11), it is straightforward to verify that this gauge-fixed action is Lorentz-invariant.

One may wonder whether this action is Hermitian in all reference frames since the transformations (11) do not preserve the Hermiticity of $A_\mu$. The answer is this: One may pick a reference frame in which $A_\mu$, $c^a$, and $b^\alpha$ are Hermitian, while $\overline{c}^A$ and $\overline{c}^\alpha$ are anti-Hermitian. In this reference frame it can be easily verified that the action of (18) is Hermitian. If the action is Hermitian in that one frame, then as a result of Lorentz invariance, it is Hermitian in all reference frames even though each field does not individually maintain its Hermiticity.

It should be noted that a slightly different choice of gauge-fixing fermion can be used to introduce a mass scale into the action. Using (1) and (2), if one adds a term of $mc^8$ into the gauge-fixing fermion, then there will be a new term in the gauge-fixed action

$$-\tfrac{1}{2}mf^{8\alpha\beta}c^\alpha c^\beta = \tfrac{1}{3}im\text{Tr}(F^8 cc) \tag{19}$$

which can be seen to be Lorentz-invariant using (11). This behavior of this "mass" term will not be explored here.

**Conclusion**

A new Lorentz-covariant gauge has been presented for SU(3). The new gauge includes a generalized axial gauge condition in the (4, 5, 6, 7) adjoint gauge directions. In order for the new gauge-fixed action to remain Lorentz-invariant, every Lorentz transformation must be accompanied by a generalized global gauge transformation. This global gauge transformation imparts an additional spin ½ to both ghosts and gauge fields in the (4, 5, 6, 7) directions. With this extra spin, the gauge fields in those directions have the wrong relationship between spin and statistics, so they cannot exist asymptotically, whereas the ghosts have the correct relationship, so they can exist asymptotically. As a result, the gauge condition asymptotically breaks SU(3) down to an SU(2)xU(1) symmetry with spin ½ ghosts.

It is tempting to compare the SU(2)xU(1) symmetry constructed in this way with the SU(2)xU(1) symmetry of the Standard Model. For example, if one introduced a left-handed triplet of leptons to interact with SU(3) and called them ("neutrino", "electron", "positron"), then after symmetry breaking, the triplet would break into a doublet and a singlet with the correct quantum numbers. The SU(3)-derived theory would have an "electroweak" mixing angle of exactly $\sin^2 \theta_W = \frac{1}{4}$ as opposed to the measured value of 0.23 [4], but differences in the theories could also affect the effective angles measured experimentally. Could the spin ½ ghosts play the role of quarks in this model? It is intriguing that the factors of 1/3 in front of the traces in the Lagrangian translate into factors of 1/3 in front of the "electric" charge interaction terms, and that BRST constraints on ghosts have the effect of "confining" ghosts into BRST-invariant combinations. It would be interesting to further explore these analogies to see how well they would hold up.

In any case, it is hoped that the present work will open a discussion around whether ghosts could acquire spin and actually be "seen" asymptotically, whether symmetry breaking could be stimulated by a gauge condition, and whether it is possible to have Lorentz-invariant gauges without derivatives.

# References

[1] M. Henneaux and C. Teitelboim, *Quantization of Gauge Systems*, Princeton University Press (Princeton, 1992).

[2] P. Ramond, *Field Theory: A Modern Primer*, Addison-Wesley (Redwood City, 1987).

[3] J. Bjorken and S. Drell, *Relativistic Quantum Fields*, McGraw-Hill (New York, 1965).

[4] The LEP Collaborations, the LEP Electroweak Working Group, the SLD Electroweak and Heavy Flavour Groups, Phys. Rep. 427 (2006) 257 [arXiv:hep-ex/0509008].